\documentclass[pra,twocolumn,superscriptaddress,amsmath,amssymb,floatfix,noshowpacs,nofootinbib]{revtex4-2} 
\usepackage{graphicx}   
\usepackage{dcolumn}    
\usepackage{bm}         
\usepackage[toc,page]{appendix}

\begin{document}

\title{\textbf{Free-space quantum interface of a single atomic tweezer array with light}}
\author{Yakov Solomons$^*$}
\affiliation{Department of Chemical \& Biological Physics, Weizmann Institute of Science, Rehovot 7610001, Israel}
\author{Roni Ben-Maimon$^*$}
\affiliation{Department of Chemical \& Biological Physics, Weizmann Institute of Science, Rehovot 7610001, Israel}
\author{Arpit Behera$^*$}
\affiliation{Department of Physics of Complex Systems, Weizmann Institute of Science, Rehovot 7610001, Israel}
\author{Ofer Firstenberg}
\affiliation{Department of Physics of Complex Systems, Weizmann Institute of Science, Rehovot 7610001, Israel}
\author{Nir Davidson}
\affiliation{Department of Physics of Complex Systems, Weizmann Institute of Science, Rehovot 7610001, Israel}
\author{Ephraim Shahmoon}
\affiliation{Department of Chemical \& Biological Physics, Weizmann Institute of Science, Rehovot 7610001, Israel}
\date{\today}

\begin{abstract}
We present a practical approach for interfacing light with a two-dimensional atomic tweezer array. Typical paraxial fields are poorly matched to the array's multi-diffraction-order radiation pattern, thus severely limiting the interface coupling efficiency. Instead, we propose to design a field mode that naturally couples to the array: it consists of a unique superposition of multiple beams corresponding to the array's diffraction orders.
This composite mode can be generated from a single Gaussian beam using standard free-space optics, including spatial light modulators and a single objective lens. For a triangular array with lattice spacing about twice the wavelength, all diffraction angles remain below $35^\circ$, making the scheme compatible with standard objectives of numerical aperture $\mathrm{NA} \le 0.7$. Our analytical theory and scattering simulations reveal that the interface efficiency $r_0$ for quantum information tasks scales favorably with the array atom number $N$: reaching $>0.99$ ($>0.9999$) for $N=149$ ($N\sim 1000$) and scaling as $1-r_0\sim 1/N$ for large $N$. The scheme is robust to optical imperfections and atomic-position errors, offering a viable path for quantum light–matter applications and state readout in current tweezer-array platforms.
\end{abstract}

\maketitle
\def\thefootnote{*}\footnotetext{These authors contributed equally to this work}

\section{Introduction}

The manipulation of quantum states of atoms and light is central to various applications in quantum optical science. Crucially, it relies on establishing an efficient interface between internal atomic states and an accessible ``target" photon mode, to enable quantum tasks such as quantum state transfer, memory and entanglement \cite{ref55,ref15,ref17,ref19,ref64}.

Such quantum interfacing to light is of particular importance for arrays of atoms trapped in optical tweezers, which have emerged as a prominent platform for quantum science and technology \cite{ref02,ref04,ref05,ref06,ref08,ref09,ref11,ref14,ref65,schlosser2023}. Nevertheless, the interface efficiency of atomic tweezer arrays is typically very limited due to the poor spatial overlap between the array's radiation pattern and paraxial light. Specifically, the lattice spacing $a$ of typical two-dimensional (2D) tweezer arrays exceeds the resonant wavelength of light $\lambda$, resulting in a radiation pattern comprised of multiple lattice diffraction orders (Fig. 1a). Meanwhile, the typical target photon mode is paraxial and hence couples only to the normal-incident zeroth diffraction order of the array. This results in an interface coupling strength $\Gamma$ to the desired target mode that is significantly smaller than the scattering loss rate $\gamma_{\mathrm{loss}}$ to higher diffraction orders uncoupled to the target mode, yielding a poor interface efficiency $r_0=\Gamma/(\Gamma+\gamma_{\mathrm{loss}})\ll 1$ \cite{Uni}.
 Solutions include either enhancing the coupling $\Gamma$ by placing the array inside a cavity \cite{YakovCavity,ref70,COV}, or inhibiting the losses $\gamma_{\mathrm{loss}}$ using destructive interference between multiple array layers \cite{MultiLayer,MultiLayerMann}. Notably, however, these solutions require costly modifications of the system inside the vacuum chamber.

\begin{figure}[!htbp]
  \centering
  \includegraphics[width=\columnwidth]{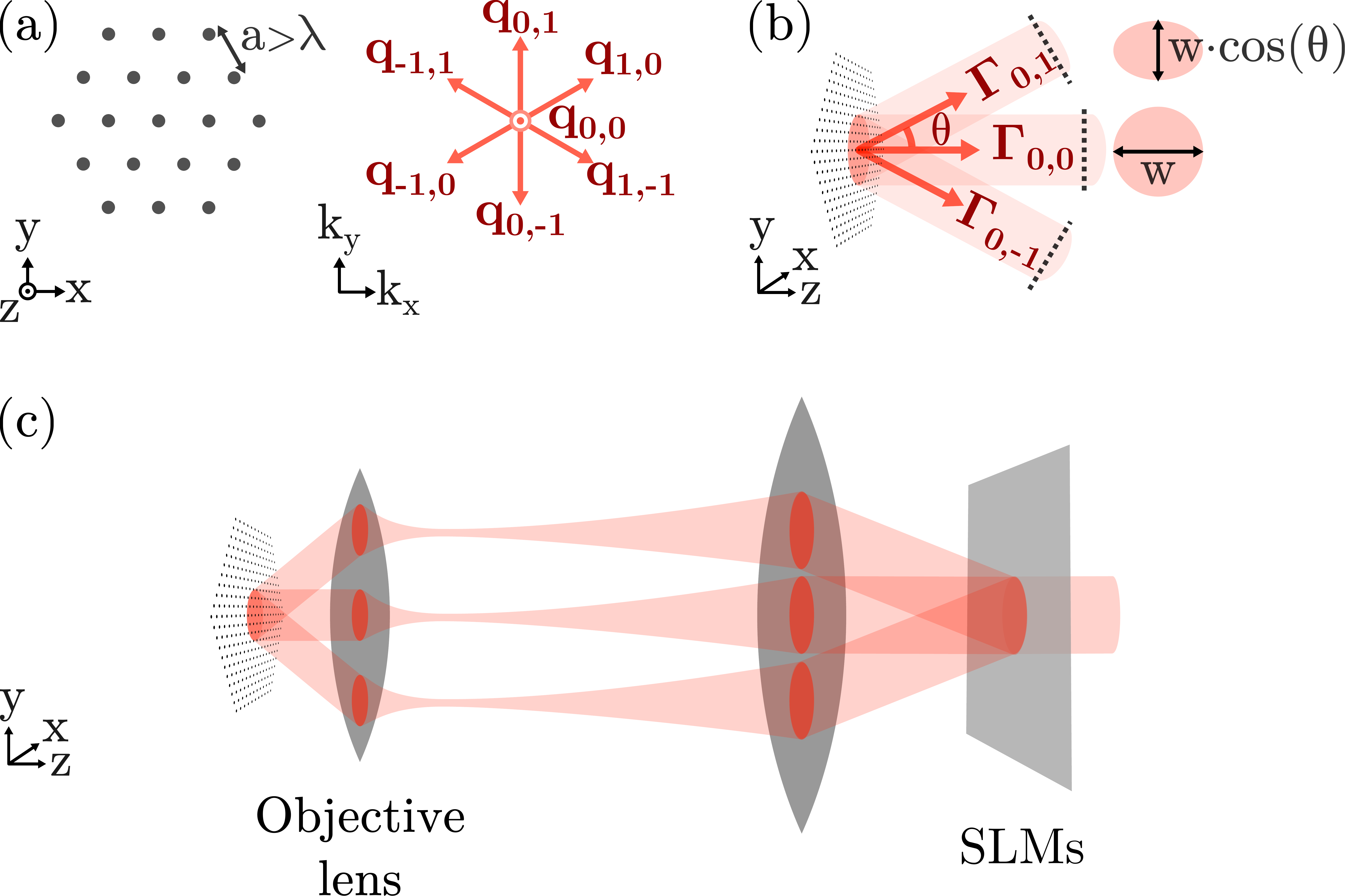}
\caption{Coupling light to a 2D tweezer array: triangular lattice. (a) For lattice spacings $a$ exceeding the wavelength $\lambda$, the uniform collective excitation of the array couples to multiple radiative diffraction orders ${\mathbf{m}}=(m_1,m_2)$ with reciprocal wavevectors $\mathbf{q}_{\mathbf{m}}$ (radiative orders for $2/\sqrt{3}<a/\lambda<2$ are shown). For coupling to a normal-incident field, only the order $\mathbf{m} = 0$ contributes (coupling rate $\Gamma_{0,0}\equiv\Gamma_{0}$), while the rest of the orders appear as losses (rates $\Gamma_{\mathbf{m}\neq 0}$). (b) The multi-beam target mode is composed of beams corresponding to all radiative diffraction orders, which now contribute to the coupling, thus yielding high coupling efficiencies. A beam component corresponding to order $\mathbf{m}$ is directed at an angle $\theta_{\mathbf{m}}$ with respect to the $z$ axis [Eq. (\ref{Gm})]. The transverse profile in the beam reference frame is an elliptical Gaussian with waists $w$ and $w\cos \theta_{\mathbf{m}}$  such that all beams form a single Gaussian of waist $w$ on the array plane. Here 3 beams out of the 7 orders from (a) are shown (with $\theta_{\mathbf{m}}\equiv \theta$ for $\mathbf{m}\neq 0$). (c) Setup for generating (collecting) the multi-beam target mode from (into) a single Gaussian beam at the input (output). Spatial light modulators (SLMs) convert the Gaussian beam into a set of beams, which are then directed through an objective lens to the array at the required angles $\theta_{\mathbf{m}}$. For a triangular lattice, a single standard objective of NA = 0.7~suffices.}\label{Fig1}
\end{figure}

In contrast, here we show that in practical situations, there exists a natural solution that relies solely on standard free-space optics (Fig.~1b,c). The idea is to design a multi-beam target mode that directly corresponds to the array's radiation pattern: the mode consists of the unique superposition of the array's diffraction orders that naturally couples to the array. In this scheme, these diffraction orders become part of the coupling $\Gamma$ instead of the loss $\gamma_{\mathrm{loss}}$, resulting in near-unity efficiencies. We identify key insights that make this solution practical. First, considering typical lattice spacings of up to a few $\lambda$, the number of radiative diffraction orders is rather small; up to $7$ diffraction orders for a triangular lattice of $a < 2\lambda$. This makes feasible the conversion and manipulation of the multi-beam mode directly from a single Gaussian beam via standard free-space optics, such as spatial light modulators (SLMs) and on-axis imaging.  Second, we show that for a proper choice of the lattice spacing, the diffraction angles can become quite low: remarkably, this allows using an optical setup with a moderate numerical aperture (NA), realized by standard objective lenses. In particular, for a triangular lattice with $a\lesssim 2\lambda$ we find the diffraction angle $\sim35^{\circ}$, corresponding to $\mathrm{NA} \approx 0.57$.

We present an analytical theory of this multi-beam quantum interface: the unique superposition of the target-mode beams and polarizations is derived and used to estimate the interface efficiency $r_0$, finding excellent agreement with $r_0$ extracted from direct numerical calculations of scattering. We study $r_0$ in practical situations, considering finite beam and array sizes, finite lens aperture, and imperfections in atomic positions. For the finite-size triangular array with lattice spacings $a\sim 1.8\lambda$, we show that a single standard objective lens with NA~=~0.7 is sufficient to achieve high interface efficiencies. Specifically, we find efficiencies of $r_0>0.99$ for $N=149$ atoms (or only dozens of atoms for NA exceeding 0.7) and $r_0>0.9999$ for $N\sim 1000$ atoms, with the favorable scaling $1/N$ of the inefficiency for large $N$. Our results thus provide a practical approach for coupling light to current mesoscopic tweezer arrays with wavelength-scale lattice spacings.

\section{Multi-beam quantum interface with a 2D array}

\subsection{Basic idea}
Consider an array of $N$ two-level atoms forming a 2D lattice on the $xy$ plane at $z=0$. The transverse ($xy$) position of an atom $\mathbf{n}=(n_1,n_2)$ (with $n_{1,2}$ integers) is given by $\mathbf{r}_{\mathbf{n}}=(n_1 a+n_2a\cos\psi,n_2 a\sin\psi)$, with $a$ being the lattice spacing and $\psi=\pi/3$ ($\psi=\pi/2$) for a triangular (square ) lattice. In order to analyze the operation of the array as a quantum light-matter interface, we first discuss an ideal infinite array and then extend the results to the realistic finite-size case.

We begin by focusing on the collective dipole given by the symmetric superposition of all atomic dipoles, $\hat{P}=\frac{1}{\sqrt{N}} \sum_{{\mathbf{n}}} \hat{\sigma}_{\mathbf{n}}$, with $\hat{\sigma}_{\mathbf{n}}$ denoting the two-level lowering operator of the dipole transition of atom $\mathbf{n}$. Classically, a collective excitation $\hat{P}$ amounts to all atoms radiating in phase, which results in radiation directed at all propagating diffraction orders of the atomic 2D lattice. Namely, the collective dipole $\hat{P}$ is coupled to all the plane waves with in-plane ($xy$) wavevectors $\mathbf{q}_{\mathbf{m}}=\frac{2\pi}{a}\left(m_{1},-m_1 \cot\psi+m_2\frac{1}{\sin\psi}\right)$  corresponding to the reciprocal lattice vectors $\mathbf{m}=(m_1,m_2)$ of the 2D lattice and that satisfy $k_{z}^{\mathbf{m}}/k=\cos\theta_{\mathbf{m}}=\sqrt{1-|\mathbf{q}_{\mathbf{m}}|^{2}/k^2}\in\text{Re}$, with $k=2\pi/\lambda$ being the incident wavenumber and $\theta_{\mathbf{m}}$ the angle at which the order $\mathbf{m}$ is directed (Fig.~1a).

Notably, for a subwavelgnth array, $a<\lambda$, only the zeroth diffraction order $\mathbf{m}=0$ is propagating ($k_{z}^{\mathbf{m}}\in\text{Re}$). Therefore, taking the normally directed order $\mathbf{m}=0$ as the target mode of a quantum interface with the array atoms yields excellent interface efficiencies, as studied before and observed in an optical lattice system \cite{ref28,ref27,Efi2017,ref29,ref30,ref31,ref32,ref37,ref38,ref40,ref18,ref41,ref42,ref43,ref44,ref45,ref46,ref47,ref26,ref48}. However, for tweezer-array platforms, where typically $a>\lambda$, additional diffraction orders $\mathbf{m}$ become propagating, to which the array radiates at the corresponding rates \cite{Uni}
\begin{eqnarray}
&&\Gamma_{\mathbf{m}}=\Gamma_{0}\frac{1-|\mathbf{q}_{\mathbf{m}}\cdot\mathbf{e}_{d}|^{2}/k^2}{\cos\theta_{\mathbf{m}}},
\quad \Gamma_0=\gamma\frac{3}{4\pi}\frac{\lambda^2}{a^2},
\nonumber \\
&&\quad \text{for} \quad  \cos\theta_{\mathbf{m}}=k_{z}^{\mathbf{m}}/k=\sqrt{1-|\mathbf{q}_{\mathbf{m}}|^{2}/k^2}\in\text{Re},
\label{Gm}
\end{eqnarray}
with $\gamma$ the spontaneous emission rate of a single atom, and $\mathbf{e}_{d}\bot \mathbf{e}_z$ the unit vector of the dipole matrix element of the atomic transition [taken below as circular polarization, $\mathbf{e}_{d}=(\mathbf{e}_{x}+i\mathbf{e}_{y})/\sqrt{2}$]. 

For an interface between a tweezer array $a>\lambda$ and the normally directed target mode, the coupling rate $\Gamma$ is given by that of the zeroth diffraction order ($\mathbf{m}=0$) $\Gamma_0$, while the radiation to higher orders $|\mathbf{m}|>0$ from Eq.~(\ref{Gm}) is seen as a loss channel, $\gamma_{\mathrm{loss}}=\sum_{\mathbf{m}\in R}\Gamma_{\mathbf{m}}$, where $R$ denotes the set of radiating diffraction orders ($k_{z}^{\mathbf{m}}\in\text{Re}$) excluding $\mathbf{m}=0$. Since these losses easily exceed the target-mode coupling $\Gamma=\Gamma_0$ , they become detrimental to the efficiency of the quantum interface \cite{Uni},
\begin{eqnarray}
r_0=\frac{\Gamma}{\Gamma+\gamma_{\text{loss}}}.
\label{r0}
\end{eqnarray}
In contrast, we propose a solution based on incorporating the higher diffraction orders into the target photon mode that one shines and detects in quantum light-matter operations. This way, the coupling rate to the higher diffraction orders, $\sum_{\mathbf{m}\in R}\Gamma_{\mathbf{m}}$, is removed from the loss $\gamma_{\text{loss}}$ into the target-mode coupling $\Gamma$, thereby establishing an efficient quantum interface, $r_0\rightarrow 1$. 

The required target mode is formed by a unique superposition of multiple beams (plane waves) corresponding to all the radiative diffraction orders $\mathbf{m}$, with superposition coefficients intuitively deduced as follows. The power impinged on a uniform array situated on the $xy$ plane, from a plane wave directed at an angle $\theta_{\mathbf{m}}$ relative to the $z$ axis, will gain a geometrical factor of $1/\cos\theta_{\mathbf{m}}$. Such a plane wave can come in two polarizations $\mu=s,p$ with unit vectors $\mathbf{e}^{\pm}_{\mathbf{m}\mu}$ perpendicular to the wavevector (with $\pm$ for $\pm z$ propagation). In turn, since the array dipoles are oriented at $\mathbf{e}_d$, we obtain another factor of $|\mathbf{e}^{\pm}_{\mathbf{m}\mu}\cdot \mathbf{e}_d^{\dag}|^2$ to the impinged power. This suggests that the optimal target field for coupling to the uniform collective dipole $\hat{P}$ is composed of the superposition of normalized plane waves at radiative diffraction orders $\mathbf{m}$ and corresponding polarization $\mu$ with superposition coefficients
\begin{eqnarray}
c^{\pm}_{\mathbf{m}\mu}=\frac{\mathbf{e}^{\pm}_{\mathbf{m}\mu}\cdot \mathbf{e}_d^{\dag}}{\sqrt{\cos\theta_{\mathbf{m}}}}.
\label{cm}
\end{eqnarray}
Notably, for a given order $\mathbf{m}$, summing the resulting impinged power over both polarizations, we obtain $\sum_{\mu=s,p}|c^{\pm}_{\mathbf{m}\mu}|^2=(1-|\mathbf{q}_{\mathbf{m}}\cdot\mathbf{e}_{d}|^{2}/k^2)/\cos\theta_{\mathbf{m}}$ in agreement with the decay rates $\Gamma_{\mathbf{m}}$ from Eq.~(\ref{Gm}).

\subsection{Formal description}
To make the above ideas more formal, we employ a generic 1D model of a quantum interface to which the array problem can be mapped. The model describes the coupling at rate $\Gamma$ between a collective dipole $\hat{P}$ and a 1D propagating target photon mode $\hat{\mathcal{E}}(z)$ as per the Heisenberg-picture equations \cite{Uni},
\begin{eqnarray}
&&\dot{\hat{P}}=\left[i\left(\delta-\Delta\right)-\frac{\Gamma+\gamma_{\text{loss}}}{2}\right]\hat{P}+i\sqrt{\Gamma}{\cal \hat{E}}_{0}(t)\left(0\right)+\hat{F}(t),
\nonumber\\
&&\hat{{\cal E}}\left(z\right)=\hat{{\cal E}}_{0}\left(z\right)+i\sqrt{\Gamma}\hat{P},
\label{EOM}
\end{eqnarray}
where $\Delta$ describes a collective shift of the dipole $\hat{P}$, $\hat{\mathcal{E}}_0$ is the input field satisfying $[\hat{\mathcal{E}}_0(t),\hat{\mathcal{E}}^{\dag}_0(t')]=\delta(t-t')$, and $\delta$ is the single-atom detuning from the central frequency of the incident field.
In addition to the target mode, the collective dipole is coupled to lossy modes at rate $\gamma_{\text{loss}}$, with corresponding quantum noise $\hat{F}$. The efficiency of the quantum interface is then universally given by $r_0$ from Eq.~(\ref{r0}), as demonstrated for various quantum tasks such as quantum memory and entanglement generation \cite{Uni,ref48,ref26}. This holds both for the linear version of the model taken here (with $[\hat{P},\hat{P}^{\dag}]=1$), and for relevant nonlinear variants. For a planar system such as a 2D atomic array, the target mode symmetrically propagates in both sides, $\hat{{\cal E}}=[\hat{{\cal E}}_+ +\hat{{\cal E}}_-]/\sqrt{2}$, and the efficiency $r_0$ is equal to the reflectivity of the array to light shined from either side, $\hat{{\cal E}}_{\pm}$. This allows to extract the efficiency of quantum tasks from classical scattering \cite{Uni,ref47} --- a property we exploit below.

For our 2D array, we begin with the many-atom Heisenberg-Langevin equations of the atomic lowering operators $\hat{\sigma}_n$ and their photon-mediated dipole-dipole interactions in the linear regime. For an infinite array we first choose the normal-incident plane wave, $(k_x,k_y)=(0,0)$, as our target mode, and obtain the Heisenberg-Langevin equation for $\hat{P}=\frac{1}{\sqrt{N}} \sum_{{\mathbf{n}}} \hat{\sigma}_{\mathbf{n}}$ in the form of (\ref{EOM}) and with $\Gamma=\Gamma_0$ and $\gamma_{\text{loss}}=\sum_{\mathbf{m}\in R}\Gamma_{\mathbf{m}}>\Gamma$, leading to a poor efficiency $r_0$ as discussed above.

Instead, consider a target mode defined by the superposition described in Eq.~(\ref{cm}) above:
\begin{eqnarray}
\hat{\mathcal{E}}_{\alpha}(z)=\sqrt{\frac{\Gamma_0}{\Gamma_{\text{tot}}}}\sum_{\mathbf{m}\in R,0}\sum_{\mu=s,p}c_{\mathbf{m}\mu}^{\alpha}\hat{\mathcal{E}}_{\mathbf{m}\mu\alpha}(z).
\label{E}
\end{eqnarray}
Here $\hat{\mathcal{E}}_{\alpha}(z)$ denotes the right ($\alpha=+$) or left ($\alpha=-$) propagating part of the target mode, $\hat{\mathcal{E}}(z)=[\hat{\mathcal{E}}_+(z)+\hat{\mathcal{E}}_-(-z)]/\sqrt{2}$, $c_{\mathbf{m}\mu}^{\alpha}$ are the coefficients from Eq.~(\ref{cm}), and $\Gamma_{\text{tot}}$ is the total radiative decay rate of the array [see Eq.~(\ref{G}) below]. The sum $\sum_{\mathbf{m}\in R,0}$ runs over all radiative diffraction orders $\mathbf{m}$ (including $\mathbf{m}=0$) with polarizations $\mu=s,p$. The respective normalized field modes,
$\hat{\mathcal{E}}_{\mathbf{m}\mu\alpha}(z)=\sqrt{\cos\theta_{\mathbf{m}}}\sqrt{\frac{c}{L}}\sum_{k_z>0}\hat{a}_{\mathbf{q}_{\mathbf{m}}k_z\mu\alpha}(t)e^{i\alpha(k_z-k_z^{\mathbf{m}})z}e^{ikct}$ ($L\rightarrow \infty$), describe 1D continua $\{k_z\}$ of normalized plane-waves $[\hat{a}_{\mathbf{q}_{\mathbf{m}}k_z\mu\alpha},\hat{a}^{\dag}_{\mathbf{q}_{\mathbf{m}}k'_z\mu\alpha}]=\delta_{k_z,k'_z}$ directed at $\theta_{\mathbf{m}}$, with inputs properly satisfying the normalization $[\hat{\mathcal{E}}_{0,\mathbf{m}\mu\alpha}(t), \hat{\mathcal{E}}^{\dag}_{0,\mathbf{m}'\mu'\alpha'}(t')]=\delta(t-t')\delta_{\mathbf{m}\mathbf{m}'}\delta_{\mu\mu'}\delta_{\alpha\alpha'}$.


\begin{figure}[t]
  \centering
  \includegraphics[width=\columnwidth]{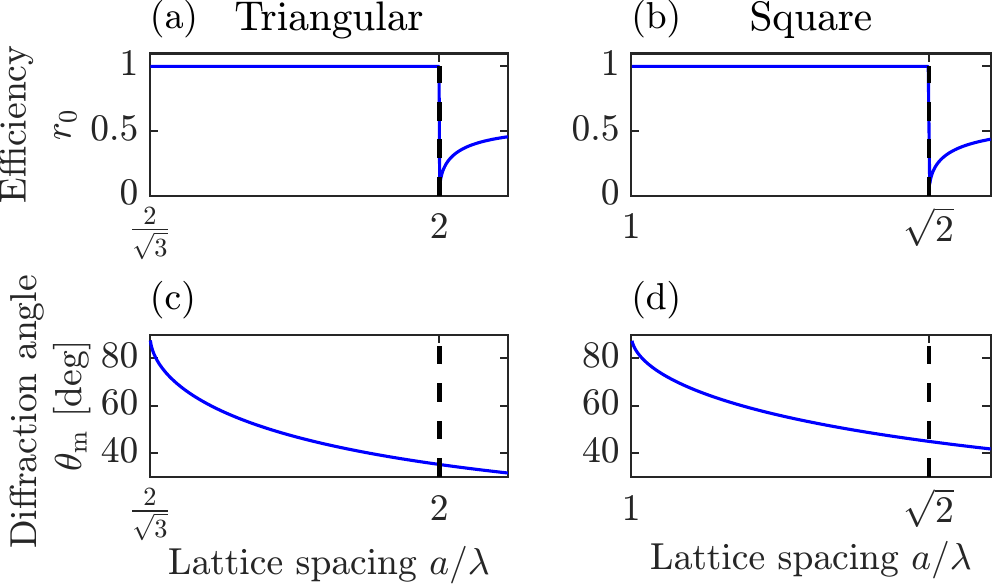}
  \caption{Infinite array theory: interface efficiency $r_{0}$ as a function of lattice spacing $a/\lambda$ for triangular (a) and square (b) arrays, considering a target mode comprising beams corresponding to the first diffraction orders $R_1$ in addition to the zeroth order $\mathbf{m} = 0$. When only this set is radiative (left of the vertical dashed line), the target mode perfectly overlaps with the array's radiation pattern yielding efficiency $r_0=1$, Eq.~(\ref{G}). For larger $a/\lambda$, more radiative orders beyond $\mathbf{m}\in\{R_1,0\}$ emerge, and the efficiency drops to $r_0=\Gamma/(\Gamma+\gamma_{\mathrm{loss}})<1$, with $\Gamma=\sum_{\mathbf{m}\in R_1,0}\Gamma_{\mathbf{m}}$, $\gamma_{\text{loss}}=\sum_{\mathbf{m}\in R,0}\Gamma_{\mathbf{m}}-\Gamma$, and $\Gamma_{\mathbf{m}}$ from Eq.~(\ref{Gm}) (text).
  (c,d) diffraction angle $\theta_{\mathbf{m}}$ of the first set of radiative orders, approaching $35^{\circ}$ ($45^{\circ}$) for the triangular (square) lattice at $a/\lambda \rightarrow 2$ ($a/\lambda \rightarrow \sqrt{2}$).}
  \label{Fig2}
\end{figure}

With the multi-beam definition (\ref{E}) for the target mode, the original many-atom Heisenberg-Langevin equations yield an equation for the collective dipole $\hat{P}=\frac{1}{\sqrt{N}} \sum_{{\mathbf{n}}} \hat{\sigma}_{\mathbf{n}}$ that again takes the form (\ref{EOM}), however, this time with (Appendix A)
\begin{eqnarray}
\Gamma=\Gamma_{\text{tot}}=\sum_{\mathbf{m}\in R,0} \Gamma_{\mathbf{m}}=\Gamma_0+\sum_{\mathbf{m}\in R}\Gamma_{\mathbf{m}}, \quad \gamma_{\text{loss}}=0.
\label{G}
\end{eqnarray}
Namely, the coupling rate to the higher diffraction orders $\sum_{\mathbf{m}\in R}\Gamma_{\mathbf{m}}$ is now removed from the loss $\gamma_{\text{loss}}$ into the coupling $\Gamma$ to the target mode. For an ideal infinite array, this leads to perfect efficiency $r_0=1$.

If the target mode does not include \emph{all} the $\mathbf{m}\neq 0$ radiative diffraction orders $R$ but only a subset $R_1\in R$, one obtains $\Gamma=\sum_{\mathbf{m}\in R_1,0}\Gamma_{\mathbf{m}}$ and
$\gamma_{\text{loss}}=\sum_{\mathbf{m}\in R,0}\Gamma_{\mathbf{m}}-\Gamma$ so that the efficiency $r_0$ drops below unity, even in the infinite array case. This is illustrated in Fig.~2, where the efficiencies of triangular and square infinite arrays are plotted as a function of the array lattice spacing $a$. In both cases, the target mode includes a single set of radiative diffraction orders beyond the zeroth order $\mathbf{m}=0$.
For the triangular array (Fig.~\ref{Fig2}a) this corresponds to the radiative orders $\mathbf{m}=\{(0,0),(\pm1,0),(0,\pm1),(\pm1,\pm1)\}$, requiring a multi-beam target mode consisting of $6$ beams (each spanned by polarizations $\mu=s,p$), in addition to the normal-incident beam $\mathbf{m}=0$. For $2/\sqrt{3}<a/\lambda<2$ these beams represent all of the existing radiative diffraction orders, yielding perfect efficiency $r_0=1$. In contrast, when $a$ is increased further, more radiative orders emerge and the efficiency drops. Perfect efficiency then requires to include the additional diffraction orders in the target mode. Similarly, for the square array (Fig.~\ref{Fig2}b), the target mode includes $4$ beams corresponding to the orders $\mathbf{m}=\{(\pm1,0),(0,\pm1)\}$, apart from the zeroth-order beam $\mathbf{m}=0$: in the range $1<a/\lambda<\sqrt{2}$, where these represent all radiative orders, we observe $r_0=1$, while for larger $a$ we obtain $r_0<1$.

\begin{figure}[t]
  \centering
  \includegraphics[width=\columnwidth]{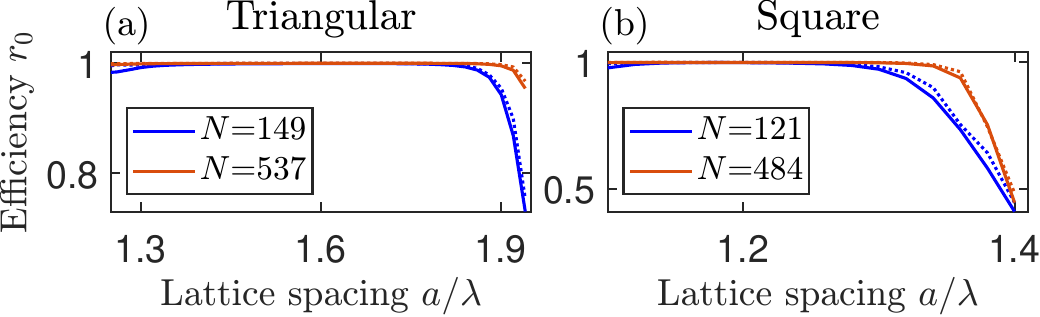}
  \caption{Interface efficiency $r_{0}$ as a function of lattice spacing $a/\lambda$ for triangular (a) and square (b) finite-size arrays (atom number $N$). In both cases, we plot the region of $a/\lambda$ where only the set of the first diffraction orders and the zeroth order $\mathbf{m} = 0$ are radiative, taking the target mode containing all corresponding finite-waist beams. $r_{0}$ is evaluated numerically from scattering calculations of reflectivity (solid lines) and theoretically from Eq.~(\ref{r0}) with (\ref{Gu}) (dotted lines). For each data point, the reflectivity is optimized with respect to the beam waist $w$ (Appendix B).}\label{Fig3}
\end{figure}

\section{Finite-size arrays}
We now show how the above ideas work in realistic finite-size arrays. For a 2D array of $N$ atoms and lattice spacing $a$, we would have to consider a multi-beam target mode whose constituent beams have a finite cross-section on the array plane that fits within the array linear dimension $L_a\sim a\sqrt{N}$. Therefore, in the superposition (\ref{E}) we now replace the plane waves directed at $\theta_{\mathbf{m}}$ with polarizations $\mu$ by corresponding Gaussian beams of finite waists. The different beams are designed to form an identical Gaussian profile $\propto e^{-(x^2+y^2)/w^2}$ on the array plane $z=0$ with the waist $w<L_a\sim \sqrt{N}a$. This means that for an oblique-incident beam, $\theta_{\mathbf{m}}\neq 0$, the normalized Gaussian profile $u_{\mathbf{m}}(x',y')$ in the beam reference frame $\{x',y',z'\}$ (with $z'$ tilted by an angle $\theta_{\mathbf{m}}$ with respect to $z$), has an ellipse shape (Fig.~\ref{Fig1}b): E.g. $u_{1,0}(x',y') = \sqrt{\frac{2}{\pi w_{x'} w_{y'}}}e^{-\frac{x'^2}{w_{x'}^2}-\frac{y'^2}{w_{y'}^2}}$ with $w_{x'}=w\cos\theta_{1,0}$ and $w_{y'}=w$ for the beam $\mathbf{m}=(1,0)$ in a square array (Appendix B).

To evaluate the interface efficiency $r_{0}$ in the finite-size case, we begin by performing numerical simulations of electromagnetic wave scattering. Considering an incident right-propagating target mode comprised of the superposition of the Gaussian beams $\propto u_{\mathbf{m}}\mathbf{e}^{+}_{\mathbf{m}\mu}$ described above with coefficients $c^{+}_{\mathbf{m}\mu}$ from Eq.~(\ref{cm}), we extract the resonant reflectivity to the same multi-beam mode (see Appendix B for details). In turn, this reflectivity is equivalent to the efficiency $r_0$ \cite{Uni}, as guaranteed by the mapping of the system to Eq. (\ref{EOM}), established in Sec.~II B.

The results of $r_0$ for finite-size  triangular and square arrays are presented in Fig.~\ref{Fig3} (solid lines) as a function of the array lattice spacing $a$ in the region where there exists a single set of radiative diffraction orders beyond $\mathbf{m}=0$ (recalling $2/\sqrt{3}<a/\lambda<2$ and $1<a/\lambda<\sqrt{2}$ for triangular and square arrays, respectively). The target mode is taken, as before, to include all the beams corresponding to these radiative diffraction orders [with $\mathbf{m}=\{(0,0),(\pm1,0),(0,\pm1),(\pm1,\pm1)\}$ and $\mathbf{m}=\{(0,0),(\pm1,0),(0,\pm1)\}$ for triangular and square arrays, respectively]; however this time with the finite-waist beams described above. We find in Fig.~\ref{Fig3} that the efficiency (reflectivity) $r_0$ is very high ($>0.99$ with $N\geqslant 121$) for values of $a/\lambda$ well within the chosen region, in agreement with the infinite-array theory. This changes at the edges of the region where $r_0$ is seen to drop.

\subsection{Competing finite-size effects}
The drop in efficiency $r_0$ at the edges of the region wherein the infinite-array theory predicts $r_0\rightarrow 1$ can be understood by analyzing finite-size effects. First, finite array size may lead to losses due to scattering of incident light off the array edges. This clearly favors a small in-plane waist $w$ of the target mode. On the other hand, as the waist $w$ gets smaller, the target-mode beams $\mathbf{m}$ contain a larger spread of transverse momenta $\mathbf{k}_{\bot}=(k_x,k_y)$ around their corresponding central momenta $\mathbf{q}_{\mathbf{m}}$. The latter has two effects that reduce the efficiency $r_0$:

\emph{(i) Diffraction effect:} while the target-mode contains all diffraction-order beams that are radiative for the uniform in-plane momentum $\mathbf{k}_{\bot}=0$, higher momenta $|\mathbf{k}_{\bot}|>0$ may radiatively couple to higher orders $\mathbf{m}$ not contained in the target mode. This is seen by the condition $k_{z}^{\mathbf{m}}(\mathbf{k}_{\bot})=\sqrt{k^2-|\mathbf{k}_{\bot}+\mathbf{q}_{\mathbf{m}}|^{2}}\in\text{Re}$, showing that for $|\mathbf{k}_{\bot}|>0$ additional radiative diffraction orders $\mathbf{m}$ may emerge, translating to scattering losses outside of the target mode. Notably, for a given $w$ that sets the maximal $|\mathbf{k}_{\bot}|\sim 2\pi/w$, such losses become more likely as $|\mathbf{q}_m| \propto \pi/a$ becomes smaller. This explains the drop of $r_0$ observed at the upper end of the $a/\lambda$ region in Fig.~\ref{Fig3}.

\emph{(ii) Dispersion effect:} in principle, different in-plane momenta $\mathbf{k}_{\bot}$ correspond to different collective dipoles $\hat{P}_{\mathbf{k}_{\bot}}$ with corresponding collective resonance shifts $\Delta_{\mathbf{k}_{\bot}}$ \cite{Efi2017}. This means that the components $\mathbf{k}_{\bot}$ of the target-mode, taken at a central resonance frequency $\delta=\Delta_{\mathbf{k}_{\bot}=0}$, are not all simultaneously resonant with the array. Since the dispersion $\Delta_{\mathbf{k}_{\bot}}$ changes very rapidly with $\mathbf{k}_{\bot}$ near the values of $a$ where new radiative diffraction emerge \cite{Efi2017}, this effect contributes to the drop of $r_0$ at the edges of the region plotted in Fig.~\ref{Fig3}.

Therefore, while the above diffraction and dispersion effects favor a narrow spread of $\mathbf{k}_{\bot}$ values and hence a large waist $w$, the scattering from the edges of a finite-size array favors a small waist. This competition leads to the emergence of an optimal value for $w$ (as used in Fig.~3; see Appendix B, Fig.~7, for details).

\subsection{Analytical description}
We can modify the analytical theory from Sec.~II to capture these finite size effects in quantitative agreement with the numerical results. To this end, we first recall that all the beams of the target mode converge to a single Gaussian profile on the array plane, so that the relevant collective dipole $\hat{P}$ becomes
\begin{eqnarray}
\hat{P}=\frac{a}{\sqrt{\eta}}\sum_{\mathbf{n}}u(\mathbf{r}_{\mathbf{n}})\hat{\sigma}_{\mathbf{n}},
\quad
u(x,y)=\sqrt{\frac{2}{\pi w^2}}e^{-\frac{x^2+y^2}{w^2}},
\label{Pu}
\end{eqnarray}
with $\eta=\text{erf}^{2}\left(L_a/\sqrt{2}w\right)$ being the overlap between the Gaussian profile and the array of linear size $L_a\sim \sqrt{N} a$, and where $[\hat{P},\hat{P}^{\dag}]=1$ ($w\gg a$). For the target mode $\hat{\mathcal{E}}$ we take the same superposition as in Eq.~(\ref{E}), except this time the 1D modes $\hat{\mathcal{E}}_{\mathbf{m}\mu\alpha}(z)$ at angles $\theta_{\mathbf{m}}$ are taken with the corresponding Gaussian profiles $u_{\mathbf{m}}(x',y')$ described above. With these definitions for $\hat{P}$ and $\hat{\mathcal{E}}$, we obtain the mapping of the original many-atom Heisenberg-Langevin equations to the 1D model Eq.~(\ref{EOM}), with the parameters (Appendix A),
\begin{eqnarray}
\Gamma=\eta \Gamma_R,
\quad
\gamma_{\text{loss}}=\Gamma'_0-\eta\Gamma_R,
\label{Gu}
\end{eqnarray}
and
\begin{eqnarray}
\Gamma_R&=&\frac{1}{\eta}\int \frac{d\mathbf{k}_{\bot}}{(2\pi)^2}|\tilde{u}(\mathbf{k}_{\bot})|^2\sum_{\mathbf{m}\in R,0} \Gamma_{\mathbf{m}}(\mathbf{k}_{\bot}),
\nonumber\\
\Gamma_0'&=&\frac{1}{\eta}\int \frac{d\mathbf{k}_{\bot}}{(2\pi)^2}|\tilde{u}(\mathbf{k}_{\bot})|^2\sum_{\mathbf{m}} \Gamma_{\mathbf{m}}(\mathbf{k}_{\bot}),
\label{GuR}
\end{eqnarray}
where $\tilde{u}(\mathbf{k}_{\bot})=a^2\sum_{\mathbf{n}}u(\mathbf{r}_{\mathbf{n}})e^{-i\mathbf{k}\cdot\mathbf{r}_{\mathbf{n}}}$ and integrations over $\mathbf{k}_{\bot}$ are performed within the Brillouin zone of the 2D array. Here $\Gamma_{\mathbf{m}}(\mathbf{k}_{\bot})$ have the same form as $\Gamma_{\mathbf{m}}$ from Eq.~(\ref{Gm}) except for the replacement $\mathbf{q}_{\mathbf{m}}\rightarrow \mathbf{q}_{\mathbf{m}}+\mathbf{k}_{\bot}$.

That is, the coupling $\Gamma$ to the target mode, given by the sum of the decay rates $\Gamma_{\mathbf{m}}(\mathbf{k}_{\bot})$ of the diffraction orders contained in the target mode (i.e. those that are radiative for $\mathbf{k}_{\bot}=0$, given by $\mathbf{m}\in \{R,0\}$), is now integrated also over the components $\mathbf{k}_{\bot}$ weighted by the Gaussian profile $|\tilde{u}(\mathbf{k}_{\bot})|^2$. A reduction factor $\eta=\text{erf}^{2}\left(\frac{L_a}{\sqrt{2}w}\right)$ also appears, accounting for the spatial overlap with a finite size array. In turn, the resulting losses $\gamma_{\text{loss}}$ are obtained by subtracting $\Gamma$ from the total decay rate $\Gamma'_0$ of $\hat{P}$, thus accounting for losses due the diffraction effect and the scattering from the edges of the array.

This theory shows excellent agreement with the direct numerics described above, as seen in Fig.~3. In particular, the drop of $r_0$ near the right edge is successfully captured.

\section{Practical implementation}
So far, we have shown that applying a multi-beam target mode can, in principle, transform a tweezer array into an efficient light–matter interface. A practical question is how such a mode can be physically realized, and, in particular, whether it can be generated and manipulated using standard free-space optics with moderate NA.

For concreteness, consider the setup illustrated in Fig.~\ref{Fig1}c: the multiple beams are generated from a normal-incident Gaussian beam using SLMs, and are then directed by an objective lens towards the required angles $\theta_{\mathbf{m}}$ on the array plane. A practical design would preferably employ a single standard-NA objective lens in front of the array. This imposes a constraint on the beam angles: they must lie within the collection cone of the objective, i.e. $\text{NA}>\sin\theta_{\mathbf{m}}$.

The lower panels of Fig.~2 show the diffraction angle $\theta_{\mathbf{m}}$ of the required beams for triangular and square arrays. Clearly, the angle becomes smaller for increasing lattice spacing $a$, favoring to work as close as possible to the right edge of the considered region. On the other hand, working too close to the right edge will result in the losses associated with the diffraction effect (Fig.~3). Triangular lattices offer a favorable compromise: for realistic lattice spacings $a/\lambda \lesssim 2$ \cite{latt1,latt2}, $\theta_{\mathbf{m}}$ can get as low as $35^{\circ}$ while maintaining a single set of radiative orders. In particular, we observe that around $a/\lambda\sim1.8$ the losses remain minimal and $\theta_{\mathbf{m}}<40^{\circ}$ (Figs. 3a and 2c), indicating compatibility with standard NA = 0.7 objectives. In the remainder of this section we therefore focus on the triangular-array configuration.
%
%
%
%
%
%
%

\subsection{Finite numerical aperture}
To test the compatibility of the triangular array with the finite span of angles supported by a realistic optical setup, we modify the numerical scattering calculation of $r_0$ to include the effect of a finite NA. To this end, given a specific value of NA, we apply a sharp low-pass spatial filter cutting all $xy$ wavevectors $|\mathbf{k}_{\bot}|>\text{NA}\cdot2\pi/\lambda$. This is performed for both the incident field on the array plane and to the scattered field just off the array. The latter, filtered reflected field is then projected on the original incident field to extract the efficiency $r_0$.

\begin{figure}[t]
  \centering
  \includegraphics[width=\columnwidth]{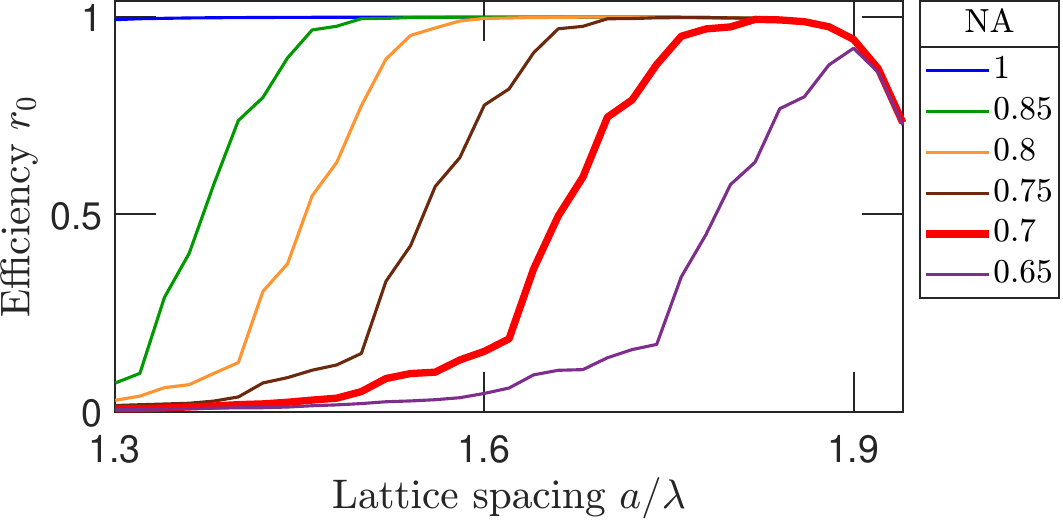}
  \caption{Interface efficiency $r_{0}$ for different numerical apertures (NAs), plotted as a function of lattice spacing $a/\lambda$ of a triangular array with $N=149$ atoms. $r_0$ is evaluated numerically from the scattering reflectivity with additional NA filtering (text), and is optimized over the waist $w$ for each NA and lattice spacing $a$.}
  \label{Fig4}
\end{figure}

The results are presented in Fig.~\ref{Fig4}, noting that the curve for NA = 1 simply reproduces that from Fig.~\ref{Fig3}a. We observe that for decreasing NA, $r_0$ drops substantially in the low $a$ region, wherein the angles $\theta_{\mathbf{m}}$ are too large to be captured by the finite NA. At the far right edge, where $a/\lambda\rightarrow 2$, we observe the same drop in $r_0$ for all values of NA, which is attributed to the diffraction effect discussed above. Between these two regions, a region of extremely high efficiency, $r_0>0.99$, can exist if the NA is not too small. Importantly, we observe that this is indeed the case for a standard NA of 0.7, with $r_0$ peaking at $a/\lambda\approx 1.82$ for the simulated atom number $N=149$.

\subsection{Scaling with atom number}
\begin{figure}[t]
  \centering
  \includegraphics[width=\columnwidth]{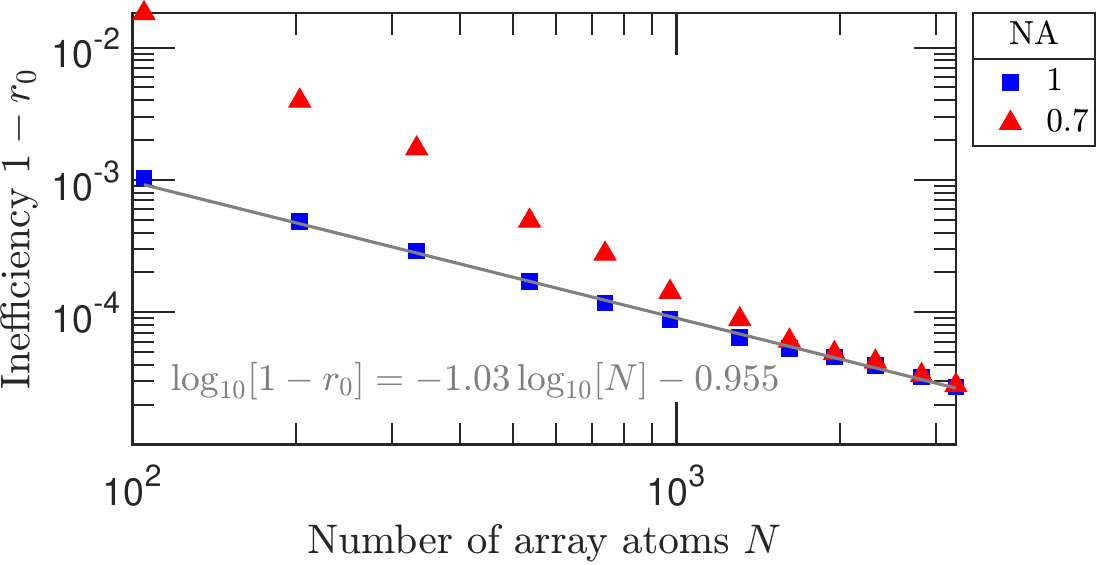}
  \caption{
Coupling inefficiency $1-r_{0}$ vs. the number of atoms $N$ in a triangular array, shown for NA = 1 (blue squares) and NA = 0.7 (red triangles). At each point, both the lattice spacing $a$ and the waist $w$ are optimized to maximize $r_{0}$ (evaluated numerically from scattering reflectivity). For NA = 1, the results are consistent with the favorable scaling $1/N$ as indicated by the fit performed for $N\geq 203$ (gray line). Similar scaling is observed for NA = 0.7 at large enough $N$.
  }\label{Fig5}
\end{figure}

We turn to study the scaling of the inefficiency $1 - r_0$ with the finite number of atoms $N$ in the triangular array.
Figure \ref{Fig5} shows the results of the numerical scattering calculation for NA = 1 and NA = 0.7. For each atom number $N$, the reflectivity is optimized over both the lattice spacing $a$ and the beam waist $w$. We observe the favorable scaling $1-r_0\sim 1/N$ of the inefficiency for NA = 1, exhibiting very low values $<10^{-2}$ for as few as tens of atoms. For NA = 0.7, the inefficiency initially decreases faster than $1/N$, while at larger $N$ it converges to the same values and scaling observed for NA~=~1: This is since larger array sizes support larger beam waists, and hence a narrower momentum-space distribution that is less affected by the angular cutoff imposed by the finite NA. Consequently, for hundreds or thousands of atoms, the inefficiencies of both NA values already become extremely small, with $1-r_0<10^{-3}$ or $1-r_0<10^{-4}$, respectively. These numbers, together with the $1/N$ scaling, suggest that the triangular-array multi-beam platform forms a practical and feasible solution for highly efficient quantum interfacing even at very moderate atom numbers.

We note that the universality of the large-$N$ behavior across the different NA cases extends also to the value of the optimal lattice spacing, which converges to a constant value $a/\lambda \approx 1.76$ for both NA cases. In contrast, for smaller atom numbers the optimal $a$ depends on $N$ and exhibits slight differences between the different NA cases (e.g. $a/\lambda \approx 1.82$ and $a/\lambda \approx 1.64$ for NA = 0.7 and NA = 1, respectively, with $N=149$).


\subsection{Imperfections in atomic positions}
Next, we consider the robustness of the setup to errors in atomic positions. To this end, we discuss two effects: the shifting of the entire array from the center of the focus of the target mode, and random errors in the positions of individual atoms.

\begin{figure}[t]
\centering
\includegraphics[width=\columnwidth]{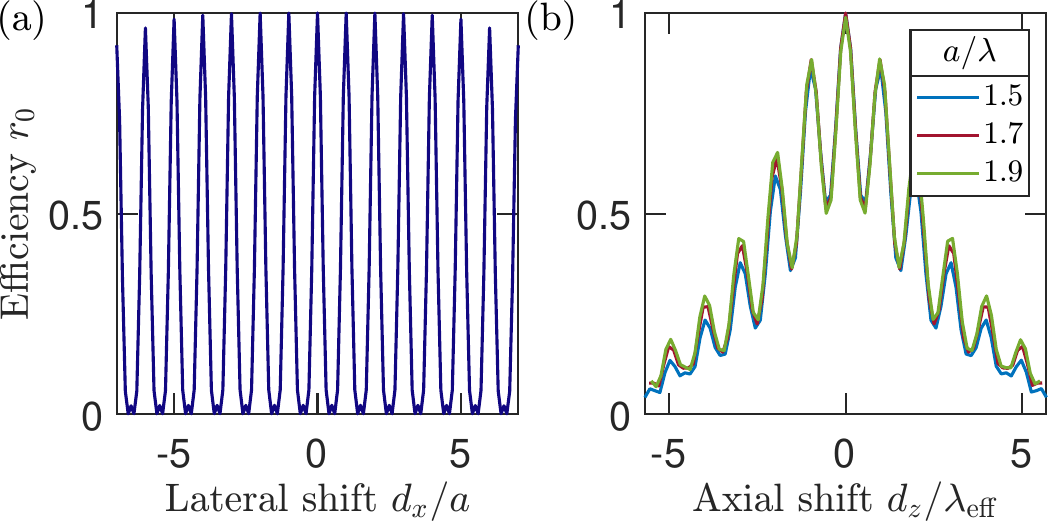}
\caption{
Dependence of efficiency $r_0$ on shifts in array position (evaluated numerically from scattering reflectivity).
(a) $r_{0}$ as a function of a lateral shift $d_{x}$ along the $x$-axis (triangular array with $N=537$ atoms, beam waist $w/L_{a} = 0.25$, and lattice spacing $a/\lambda=1.76$). (b) $r_{0}$ as a function of an axial shift $d_{z}$ along the optical axis for different lattice spacing $a/\lambda$. The horizontal axis is rescaled by the corresponding beating period $\lambda_{\text{eff}}=2\pi/(k-k_z^{\mathbf{m}})$ [same parameters as in (a), for various $a/\lambda$].
}\label{Fig6}
\end{figure}

Starting with the former, we consider the effect of a lateral shift $x=d_x$ about the center of the focus on the $xy$ plane (at $x=y=0$). The results of the efficiency $r_0$ as a function of the shift $d_x$ obtained from the numerical scattering calculation are shown in Fig.~\ref{Fig6}a. We see that $r_0$ oscillates at the lattice period $a$. This results from the fact that the superposition of all target-mode beams forms the radiative part of the reciprocal lattice, whose corresponding real-space image is the lattice itself (within diffraction-limit resolution). This effect holds for shifts $d_x$ up to the array's linear size reduced by the beam width,  i.e. $\sim \sqrt{N}a-w$, and can in fact be used as an alignment tool for the setup.

Next, we study the robustness of the efficiency $r_0$ to shifts $z=d_z$ in the array position along the optical axis, away from the focal point $z=0$. The numerical results for $r_0$ as a function of the shift $d_z$ are plotted in Fig.~\ref{Fig6}b.
We observe oscillations of $r_0$ at the beating period $\lambda_{\text{eff}}=2\pi/(k-k_z^{\mathbf{m}})$ at which the zeroth diffraction order (with wavenumber $k$ along $z$) and the higher radiative orders $\mathbf{m}\neq 0$ (with wavenumber $k_z^{\mathbf{m}}$ identical to all in our case) rephase together to form the correct unique superposition [Eq.~(\ref{cm})]. Notably, different lattice spacings $a$ exhibit different values of $k_z^{\mathbf{m}}$ and hence different $\lambda_{\text{eff}}$: the plots for different $a$ then approximately collapse onto each other when $d_z$ is rescaled by $\lambda_{\text{eff}}$. In addition, the decreasing envelope decays with a lengthscale $l\sim w/\tan\theta_{\mathbf{m}}$ within which the normal-incident ($\mathbf{m}=0$) and angled  beams ($\mathbf{m}\neq0$) spatially overlap to form the desired superposition.

Finally, consider random position errors $\delta r$ of individual atoms around their ideal lattice sites. These uncorrelated displacements introduce random phases into the scattered field which, to lowest-order in $\delta r/\lambda$, are equivalent to an individual-atom loss term scaling as $\gamma_{\text{loss}}\sim \Gamma_0 (\delta r/\lambda)^2$ \cite{Uni,Efi2017}. This adds a contribution of $\sim(\Gamma_0/\Gamma) (\delta r/\lambda)^2\sim  (\delta r/a)^2$ to the inefficiency $1-r_0$ (noting $\Gamma\sim \Gamma_0+\sum_{\mathbf{m}\in R}\Gamma_{\mathbf{m}}\sim \gamma\sim(a/\lambda)^2\Gamma_0$ \cite{YakovCavity}).

\section{Discussion}
In this work, we presented a method for efficiently coupling atomic tweezer arrays to propagating light, addressing the common challenge posed by array lattice spacings which exceed the wavelength of light. Considering the emergence of tweezer arrays as a leading platform for quantum information processing, such efficient light–matter interfaces can have significant impact on a wide range of applications.

For instance, the rate and fidelity of quantum-state readout in tweezer-array qubits are often limited by their weak fluorescence collection. This challenge becomes especially rlevant for quantum computation schemes involving mid-circuit measurements \cite{ref77,ref78,ref69,ref70,ref71}. Typical free-space detection relies on single-atom readout with efficiencies of only a few percent \cite{ref66,ref67,ref68}. In contrast, our results show that a compact array of just $\sim 60$ atoms, addressed with a NA = 0.7 objective, can already achieve an efficiency of $r_0\sim 0.9$. This enhancement relies on the efficient multi-beam coupling to a \emph{collective} atomic excitation in the array, and could be harnessed in two ways: (i) if the quantum information is encoded and manipulated directly at the collective excitation, or (ii) if such a "patch array" is used as an optical antenna to extract or mediate quantum information from other qubits in a larger array \cite{ref32,ref74}.

Another direction is the use of tweezer arrays as light-matter interfaces for quantum information tasks, such as quantum memories. Arrays of hundreds or thousands of atoms are already experimentally accessible \cite{ref72,ref73}; for such sizes we recall our predictions of interface efficiencies of $0.999$ to $0.9999$, well-suited for high-fidelity applications. These interfaces also enable studies of quantum nonlinear optics using Rydberg levels \cite{ref75,ref76}, without requiring large optical depths \cite{ref48,ref45,ref41}. Notably, the quantum memory implementation requires illuminating the array with the multi-beam target mode from both sides \cite{ref29,Uni}, which can be achieved using beam splitters and other standard optical elements. In contrast, generating photonic correlations via Rydberg interactions requires only single-sided illumination \cite{ref18,ref41,ref45,ref48}, as in the configuration shown in Fig.~1.

\begin{acknowledgments}
We acknowledge financial support from the Israel Science Foundation (ISF), the Directorate for Defense Research and Development (DDR\&D), the Minerva
Stiftung with funding from the Federal German Ministry for Education and Research, the US-Israel Binational Science Foundation (BSF) and US National Science Foundation (NSF), the Center for New Scientists at the Weizmann Institute of Science, the Council for Higher Education (Israel), the Helmsley Charitable Trust, and the Estate of Louise Yasgour. This research is made possible in part by the historic generosity of the Harold Perlman Family.
\end{acknowledgments}

\appendix
\section{Analytical theory}
Starting from the full Hamiltonian of $N$ atoms coupled to quantized field modes in free space, and applying standard Born-Markov type approximation, one obtains the Heisenberg-Langevin equations for the atomic and field operators in a rotated frame around $kc$ and in the linear regime (number of excitations $\ll$ number of atoms) \cite{Uni}
\begin{eqnarray}
&&\frac{d\hat{\sigma}_n}{dt}=i\delta \hat{\sigma}_n+i\frac{d}{\hbar}\hat{E}_0(\mathbf{r}_n)+i\frac{3}{2}\gamma\lambda \sum_m G(\mathbf{r}_n-\mathbf{r}_m)\hat{\sigma}_m,
\nonumber\\
&&\hat{\mathbf{E}}(\mathbf{r})=\hat{\mathbf{E}}_0(\mathbf{r})+\frac{k^2 d}{\varepsilon_0}\sum_n \overline{\overline{G}}(\mathbf{r}-\mathbf{r}_n)\cdot \mathbf{e}_d \hat{\sigma}_n.
\label{A1}
\end{eqnarray}
Here $d$ is the atomic dipole matrix element and $\overline{\overline{G}}(\mathbf{r})$ is the dyadic Green's tensor of the field, while $G=\mathbf{e}^{\dag}_d\cdot \overline{\overline{G}}\cdot \mathbf{e}_d$ and $\hat{E}(\mathbf{r})=\mathbf{e}^{\dag}_d\cdot \hat{\mathbf{E}}(\mathbf{r})e^{ikct}$ are projections of the Green’s tensor and the photon field operator onto the dipole orientation $\mathbf{e}_d$. The field operator in free space is given by
\begin{eqnarray}
\hat{\mathbf{E}}(\mathbf{r})=\sum_{\mathbf{k}_{\bot}}\sum_{k_z}\sum_{\mu=s,p}
\sqrt{\frac{\hbar\omega_{\mathbf{k}_{\bot}k_z}}{2\varepsilon_0 A L}}e^{i (\mathbf{k}_{\bot}+k_z \mathbf{e}_z)\cdot \mathbf{r}} \mathbf{e}_{\mathbf{k}_{\bot}k_z\mu} \hat{a}_{\mathbf{k}_{\bot}k_z\mu},
\nonumber\\
\label{A2}
\end{eqnarray}
with mode frequencies $\omega_{\mathbf{k}_{\bot}k_z}=c\sqrt{|\mathbf{k}_{\bot}|^2+k_z^2}$ and quantization volume $A L\rightarrow \infty$, while the ``input" field $\hat{\mathbf{E}}_0$ is defined in the same way using the replacement $\hat{a}_{\mathbf{k}_{\bot}k_z\mu}(t) \rightarrow \hat{a}_{\mathbf{k}_{\bot}k_z\mu}(0) e^{- i\omega_{\mathbf{k}_{\bot}k_z}t}$.

For the infinite array, we obtain the equations for $\hat{P}$ and $\hat{\mathcal{E}}$ from (\ref{EOM}) with the parameters from (\ref{G}) by, respectively, summing over the atom-array variables $\hat{\sigma}_n$ and projecting the field with the multi-beam target mode superposition of plane waves $(\mathbf{q}_{\mathbf{m}},k_z^{\mathbf{m}})$ with polarizations $\mathbf{e}^{\pm}_{\mathbf{m}\mu}= \mathbf{e}_{\mathbf{q}_{\mathbf{m}},\pm k_z^{\mathbf{m}},\mu}$ described by Eqs. (\ref{cm}) and (\ref{E}) (also using the finite bandwidth of fields in $k_z$ with respect to $k$ in the Born-Markov approximation).

For the finite-size case, we span the space of $N$ atomic positions using the orthonormal basis set $v_{ln}$ ($l=0,...,N-1)$, defining the collective atomic operators $\hat{P}_l=\sum_n v^{\ast}_{ln} \hat{\sigma}_n$. We choose the mode $l=0$ to be the Gaussian from Eq.~(\ref{Pu}), $\hat{P}_{l=0}\equiv \hat{P}$, and obtain from Eq.~(\ref{A1}) the dynamical equation
\begin{eqnarray}
&&\frac{d\hat{P}}{dt}=\left[i(\delta-\Delta')-\frac{\Gamma'_0}{2}\right]\hat{P}+i\frac{d}{\hbar}\hat{E}_0-\sum_{l\neq 0} D_{0l}\hat{P}_l,
\nonumber\\
&& D_{ll'}=-i\frac{3}{2}\gamma\lambda \sum_n\sum_m v_{ln}^{\ast} G(\mathbf{r}_n-\mathbf{r}_m)v_{l'm},
\label{A3}
\end{eqnarray}
with $\Gamma'_0/2+i\Delta'\equiv D_{00}$ and $\hat{E}_0=\sum_n v_{0n}^{\ast}\hat{E}_0(\mathbf{r}_n)$. The last term in the equation for $\hat{P}$ describes mixing with other collective modes $\hat{P}_l$ via the photon-mediated dipole-dipole coupling. For large enough arrays, where $\hat{P}$ becomes an approximate dipole eigenmode \cite{Uni}, we can neglect this term. We then define the target mode operator in analogy to that of Eq.~ (\ref{E}) with the following replacements. First, the normalized 1D field modes directed at transverse momenta $\mathbf{q}_{\mathbf{m}}$ are replaced by those weighted with the Gaussian profile $\tilde{u}(\mathbf{k}_{\bot})=a^2\sum_{\mathbf{n}}u(\mathbf{r}_{\mathbf{n}})e^{-i\mathbf{k}\cdot\mathbf{r}_{\mathbf{n}}}$,
\begin{eqnarray}
\hat{\mathcal{E}}_{\mathbf{m}\mu\alpha}(z)&=&\sqrt{B_{\mathbf{m}}}\sqrt{\frac{c}{L}}
\frac{1}{\sqrt{A}}\sum_{\mathbf{k}_{\bot}}\tilde{u}^{\ast}(\mathbf{k}_{\bot})
\nonumber\\
&&\times \sum_{k_z>0}\hat{a}_{\mathbf{q}_{\mathbf{m}}+\mathbf{k}_{\bot},k_z\mu\alpha}e^{i\alpha(k_z-k_z^{\mathbf{m}})z}e^{ikct},
\label{A4}
\end{eqnarray}
noting that $\cos\theta_{\mathbf{m}}$ is now replaced by
\begin{eqnarray}
B_{\mathbf{m}}=\left[\int \frac{d\mathbf{k}_{\bot}}{(2\pi)^2}\frac{\left|\tilde{u}(\mathbf{k}_{\bot})\right|^2}
{\sqrt{1-(\mathbf{q}_{\mathbf{m}}+\mathbf{k}_{\bot})^2/k^2}}\right]^{-1}.
\label{A5}
\end{eqnarray}
Second, in the definitions of the coefficients $c^{\pm}_{\mathbf{m}\mu}$ from Eq.~(\ref{cm}), we again replace $\cos\theta_{\mathbf{m}}$ with $B_{\mathbf{m}}$ from above. Using these modified coefficients $c^{\pm}_{\mathbf{m}\mu}$ and modes $\hat{\mathcal{E}}_{\mathbf{m}\mu\alpha}(z)$ in the definition of the target mode Eq.~(\ref{E}), we are able to recast Eq.~(\ref{A3}) in the form of Eqs. (\ref{EOM}), with the parameters from Eq.~(\ref{Gu}).

\section{Numerical calculations}
To evaluate the efficiency $r_{0}$ of the finite-size array numerically, we simulate the classical scattering of a multi-beam target mode off the atomic array. Specifically, we compute the scattering of an incident target mode constructed as a superposition of Gaussian beams, each corresponding to a radiative diffraction order $\mathbf{m}$, with appropriately chosen direction, waist, and polarization, as described in Eq.~ (\ref{E}). Projecting the back-scattered part of the field onto the same target mode, we then obtain the reflectivity for this mode, which is identified as the interface efficiency $r_0$ (as generally shown in \cite{Uni}).

The simulation is an adaptation of the method described in \cite{Efi2017} to the multi-beam case. To construct the multi-beam target mode, we begin by characterizing the polarization and spatial structure of each diffraction-order beam $\mathbf{m}$. The beam associated with diffraction order $\mathbf{m}$ propagates in a well-defined direction relative to the lab frame, as illustrated in Fig.~\ref{Fig1}. The direction of propagation is specified by two angles: the diffraction angle $\theta_{\mathbf{m}}=\arcsin\left(\left|\mathbf{q}_{\mathbf{m}}\right|/k\right)$, which sets the angle between the beam and the optical z-axis, and the azimuthal angle $\phi_{\mathbf{m}}=\arctan\left(q_{\mathbf{m}}^{y}/q_{\mathbf{m}}^{x}\right)$, which determines the in-plane orientation of the beam.

To describe the spatial profile of each beam on the array plane, we define a local beam reference frame $\left\{ x'_{\mathbf{m}},y'_{\mathbf{m}},z'_{\mathbf{m}}\right\}$, where the $z'_{\mathbf{m}}$-axis is aligned with the beam's propagation direction. The transformation from the lab-frame coordinates $\boldsymbol{r}=\left(x,y,z\right)^{T}$ to the beam-frame coordinates $\boldsymbol{r}'_{\mathbf{m}}=\left(x'_{\mathbf{m}},y'_{\mathbf{m}},z'_{\mathbf{m}}\right)^{T}$ is given by
\begin{eqnarray}
\boldsymbol{r}'_{\mathbf{m}}&=\mathcal{R}_{y}\left(-\theta_{\mathbf{m}}\right)\mathcal{R}_{z}\left(-\phi_{\mathbf{m}}\right)\boldsymbol{r},
\label{frames_trans}
\end{eqnarray}
with $\mathcal{R}_i$ denoting a rotation matrix around the $i\in\{x,y,z\}$ axis.
In this beam-frame, the beam exhibits an elliptical Gaussian profile, reflecting the fact that it strikes the array at an angle $\theta_{\mathbf{m}}$, causing its originally circular waist to appear as an ellipse in the beam’s local frame. As illustrated in Fig.~\ref{Fig1}b, the circular waist of radius $w_{0}$ in the lab frame appears compressed along one direction in the beam frame. To describe this geometry, we choose the in-plane axes $x'_{\mathbf{m}}$ and $y'_{\mathbf{m}}$ such that $y'_{\mathbf{m}}$ lies entirely within the array plane and preserves the original waist, i.e.,$w_{0,y'_{\mathbf{m}}}=w$. In contrast, $x'_{\mathbf{m}}$ which lies in the beam’s plane of incidence and is tilted with respect to the array, experiences a compression of the waist to $w_{0,x'_{\mathbf{m}}}=w\cos\theta_{\mathbf{m}}$. This choice ensures that the projected beam footprint on the array remains circular.

The field amplitude of the elliptical Gaussian beam in the beam frame takes the form $e^{-ikz'_{\mathbf{m}}}f\left(x'_{\mathbf{m}},z'_{\mathbf{m}}\right)f\left(y'_{\mathbf{m}},z'_{\mathbf{m}}\right)$, where $f\left(\xi,z\right)$ is the normalized one-dimensional profile of a Gaussian beam given by
\begin{eqnarray}
f\left(\xi,z\right)=\sqrt{\sqrt{\frac{2}{\pi}}\frac{w_{0,\xi}}{w_{\xi}\left(z\right)}}e^{-\left[\frac{\xi}{w_{\xi}\left(z\right)}\right]^{2}-ik\frac{\xi^{2}}{2R_{\xi}\left(z\right)}+i\frac{\psi_{\xi}\left(z\right)}{2}}.
\label{um_1Dprofile}
\end{eqnarray}
Here, the beam parameters are defined as
\begin{eqnarray}
z_{R,\xi} &= \frac{\pi w_{0,\xi}^{2}}{\lambda},\quad
w_{\xi}\left(z\right) = w_{0,\xi} \sqrt{1+\left(\frac{z}{z_{R,\xi}}\right)^{2}} \nonumber \\
R_{\xi}\left(z\right) &= z\left[1+\left(\frac{z_{R,\xi}}{z}\right)^{2}\right],\quad
\psi_{\xi} = \arctan\left(\frac{z}{z_{R,\xi}}\right).
\label{beam_param}
\end{eqnarray}
The product $f\left(x'_{\mathbf{m}},z'_{\mathbf{m}}\right)f\left(y'_{\mathbf{m}},z'_{\mathbf{m}}\right)$ defines the beam's spatial profile $u_{\mathbf{m}}\left(x'_{\mathbf{m}},y'_{\mathbf{m}},z'_{\mathbf{m}}\right)$.
 The polarization directions for each diffraction-order beam are set by defining the beam-frame unit vectors $\boldsymbol{e}_{x'_{\mathbf{m}}}$ and $\boldsymbol{e}_{y'_{\mathbf{m}}}$ to align with $\boldsymbol{e}_{\mathbf{m}p}^{+}$ and $\boldsymbol{e}_{\mathbf{m}s}^{+}$ polarization directions, respectively. Transforming these vectors from the beam frame back to the lab frame yields the polarization vectors
\begin{eqnarray}
\boldsymbol{e}_{\mathbf{m}p}^{+}=\left(\begin{array}{c}
\cos\theta_{\mathbf{m}}\cos\phi_{\mathbf{m}}\\
\cos\theta_{\mathbf{m}}\sin\phi_{\mathbf{m}}\\
-\sin\theta_{\mathbf{m}}
\end{array}\right),\ \boldsymbol{e}_{\mathbf{m}s}^{+}=\left(\begin{array}{c}
-\sin\phi_{\mathbf{m}}\\
\cos\phi_{\mathbf{m}}\\
0
\end{array}\right).
\label{polarization}
\end{eqnarray}

\begin{figure}[t]
\centering
\includegraphics[width=\columnwidth]{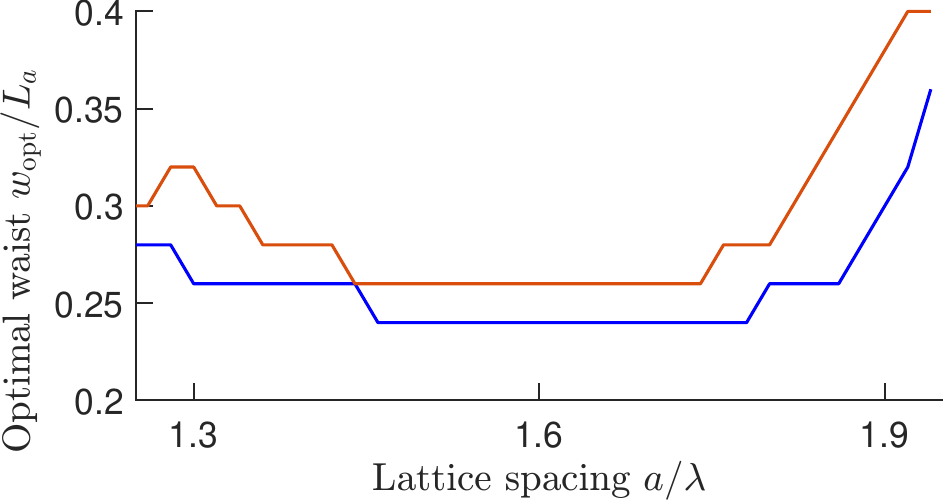}
\caption{Optimal beam waist that maximizes the efficiency (reflectivity) of the triangular array setups in Fig.~\ref{Fig3} as a function of lattice spacing $a$ (blue and red curves for $N=149$ and $N=537$, respectively).}\label{Fig7}
\end{figure}

The total incident field, defined in the lab frame and describing a right-propagating multi-beam mode composed of radiative diffraction orders $\mathbf{m}$, is expressed using beam-frame coordinates as:
\begin{align}
\boldsymbol{E}(\boldsymbol{r})
&= \sqrt{\frac{\Gamma_{0}}{\Gamma_{\text{tot}}}}
\sum_{\mathbf{m}\in R} \sum_{\mu=s,p}
\frac{\boldsymbol{e}_{\mathbf{m}\mu} \cdot \boldsymbol{e}_{d}^{\dagger}}{\sqrt{\cos\theta_{\mathbf{m}}}} \nonumber \\
&\quad \times \left[\boldsymbol{e}_{\mathbf{m}\mu}^{\dagger} e^{-ikz'_{\mathbf{m}}}
f(x'_{\mathbf{m}}, z'_{\mathbf{m}})
f(y'_{\mathbf{m}}, z'_{\mathbf{m}})\right]
\label{E_num}
\end{align}
This constructed multi-beam field serves as the incident mode in our numerical simulations, from which we compute the reflectivity $r_{0}$.

The numerical results plotted in Fig.s 3-5 are obtained with the above procedure while also optimizing the waist at each data point to maximize $r_0$. Figure~\ref{Fig7} presents the results for the optimal waist we obtained and used in Fig.~\ref{Fig3} for $r_0$ of a triangular array as a function of the lattice spacing $a$. We observe that for most values of $a$, the optimal waist settles at about $0.25$ of the array linear size $L_a\sim \sqrt{N} a$. This value manifests the balance between the dispersion effect and the effect of scattering from the array edges, favoring larger and smaller waists, respectively (see Sec. III A). Interestingly, this optimal value is consistent with those found for tweezer-array interfaces with a single-beam target mode \cite{YakovCavity,MultiLayer}.
For lattice spacings $a$ close to the upper edge $a/\lambda =2$ of the considered region, the diffraction effect discussed in Sec. III A becomes significant. Since the latter again favors larger waists to avoid losses to the next diffraction orders, the optimal waist increases as $a/\lambda$ approaches $2$. A weaker effect occurs at the lower edge of the region near $a/\lambda=2/\sqrt{3}$: there, we also identify a slight increase of the optimal value attributed to the increasing significance of the dispersion effect near the edges (Sec. III A).



%


\begin{thebibliography}{10}

\bibitem{ref55}
Klemens Hammerer, Anders~S S{\o}rensen, and Eugene~S Polzik.
\newblock Quantum interface between light and atomic ensembles.
\newblock {\em Reviews of Modern Physics}, 82(2):1041, 2010.

\bibitem{ref15}
H~Jeff Kimble.
\newblock The quantum internet.
\newblock {\em Nature}, 453(7198):1023--1030, 2008.

\bibitem{ref17}
TE~Northup and R~Blatt.
\newblock Quantum information transfer using photons.
\newblock {\em Nature photonics}, 8(5):356--363, 2014.

\bibitem{ref19}
Darrick~E Chang, Vladan Vuleti{\'c}, and Mikhail~D Lukin.
\newblock Quantum nonlinear optics—photon by photon.
\newblock {\em Nature Photonics}, 8(9):685--694, 2014.

\bibitem{ref64}
DE~Chang, JS~Douglas, Alejandro Gonz{\'a}lez-Tudela, C-L Hung, and HJ~Kimble.
\newblock Colloquium: Quantum matter built from nanoscopic lattices of atoms
  and photons.
\newblock {\em Reviews of Modern Physics}, 90(3):031002, 2018.

\bibitem{ref02}
Brian~J Lester, Niclas Luick, Adam~M Kaufman, Collin~M Reynolds, and Cindy~A
  Regal.
\newblock Rapid production of uniformly filled arrays of neutral atoms.
\newblock {\em Physical review letters}, 115(7):073003, 2015.

\bibitem{ref04}
Daniel Barredo, Sylvain De~L{\'e}s{\'e}leuc, Vincent Lienhard, Thierry Lahaye,
  and Antoine Browaeys.
\newblock An atom-by-atom assembler of defect-free arbitrary two-dimensional
  atomic arrays.
\newblock {\em Science}, 354(6315):1021--1023, 2016.

\bibitem{ref05}
Manuel Endres, Hannes Bernien, Alexander Keesling, Harry Levine, Eric~R
  Anschuetz, Alexandre Krajenbrink, Crystal Senko, Vladan Vuletic, Markus
  Greiner, and Mikhail~D Lukin.
\newblock Atom-by-atom assembly of defect-free one-dimensional cold atom
  arrays.
\newblock {\em Science}, 354(6315):1024--1027, 2016.

\bibitem{ref06}
Daniel Barredo, Vincent Lienhard, Sylvain De~Leseleuc, Thierry Lahaye, and
  Antoine Browaeys.
\newblock Synthetic three-dimensional atomic structures assembled atom by atom.
\newblock {\em Nature}, 561(7721):79--82, 2018.

\bibitem{ref08}
Antoine Browaeys and Thierry Lahaye.
\newblock Many-body physics with individually controlled rydberg atoms.
\newblock {\em Nature Physics}, 16(2):132--142, 2020.

\bibitem{ref09}
Adam~M Kaufman and Kang-Kuen Ni.
\newblock Quantum science with optical tweezer arrays of ultracold atoms and
  molecules.
\newblock {\em Nature Physics}, 17(12):1324--1333, 2021.

\bibitem{ref11}
Hannes Bernien, Sylvain Schwartz, Alexander Keesling, Harry Levine, Ahmed
  Omran, Hannes Pichler, Soonwon Choi, Alexander~S Zibrov, Manuel Endres,
  Markus Greiner, et~al.
\newblock Probing many-body dynamics on a 51-atom quantum simulator.
\newblock {\em Nature}, 551(7682):579--584, 2017.

\bibitem{ref14}
Ivaylo~S Madjarov, Jacob~P Covey, Adam~L Shaw, Joonhee Choi, Anant Kale,
  Alexandre Cooper, Hannes Pichler, Vladimir Schkolnik, Jason~R Williams, and
  Manuel Endres.
\newblock High-fidelity entanglement and detection of alkaline-earth rydberg
  atoms.
\newblock {\em Nature Physics}, 16(8):857--861, 2020.

\bibitem{ref65}
Shuo Ma, Alex~P Burgers, Genyue Liu, Jack Wilson, Bichen Zhang, and Jeff~D
  Thompson.
\newblock Universal gate operations on nuclear spin qubits in an optical
  tweezer array of yb 171 atoms.
\newblock {\em Physical Review X}, 12(2):021028, 2022.

\bibitem{schlosser2023}
Malte Schlosser, Sascha Tichelmann, Dominik Sch{\"a}ffner, Daniel~Ohl de~Mello,
  Moritz Hambach, Jan Sch{\"u}tz, and Gerhard Birkl.
\newblock Scalable multilayer architecture of assembled single-atom qubit
  arrays in a three-dimensional talbot tweezer lattice.
\newblock {\em Physical review letters}, 130(18):180601, 2023.

\bibitem{Uni}
Yakov Solomons, Roni Ben-Maimon, and Ephraim Shahmoon.
\newblock Universal approach for quantum interfaces with atomic arrays.
\newblock {\em PRX Quantum}, 5(2):020329, 2024.

\bibitem{YakovCavity}
Yakov Solomons, Inbar Shani, Ofer Firstenberg, Nir Davidson, and Ephraim
  Shahmoon.
\newblock Efficient coupling of light to an atomic tweezer array in a cavity.
\newblock {\em Physical Review Research}, 6(4):L042070, 2024.

\bibitem{ref70}
Beili Hu, Josiah Sinclair, Edita Bytyqi, Michelle Chong, Alyssa Rudelis, Joshua
  Ramette, Zachary Vendeiro, and Vladan Vuleti{\'c}.
\newblock Site-selective cavity readout and classical error correction of a
  5-bit atomic register.
\newblock {\em Physical Review Letters}, 134(12):120801, 2025.

\bibitem{COV}
William Huie, Shankar~G Menon, Hannes Bernien, and Jacob~P Covey.
\newblock Multiplexed telecommunication-band quantum networking with atom
  arrays in optical cavities.
\newblock {\em Physical Review Research}, 3(4):043154, 2021.

\bibitem{MultiLayer}
Roni Ben-Maimon, Yakov Solomons, Nir Davidson, Ofer Firstenberg, and Ephraim
  Shahmoon.
\newblock Quantum interfaces with multilayered superwavelength atomic arrays.
\newblock {\em Physical Review Letters}, 135(3):033601, 2025.

\bibitem{MultiLayerMann}
Charlie-Ray Mann, Francesco Andreoli, Vladimir Protsenko, Zala
  Lenar{\v{c}}i{\v{c}}, and Darrick Chang.
\newblock Selective radiance in super-wavelength atomic arrays.
\newblock {\em arXiv preprint arXiv:2402.06439}, 2024.

\bibitem{ref28}
G~Facchinetti, Stewart~D Jenkins, and Janne Ruostekoski.
\newblock Storing light with subradiant correlations in arrays of atoms.
\newblock {\em Physical review letters}, 117(24):243601, 2016.

\bibitem{ref27}
Robert~J Bettles, Simon~A Gardiner, and Charles~S Adams.
\newblock Enhanced optical cross section via collective coupling of atomic
  dipoles in a 2d array.
\newblock {\em Physical review letters}, 116(10):103602, 2016.

\bibitem{Efi2017}
Ephraim Shahmoon, Dominik~S Wild, Mikhail~D Lukin, and Susanne~F Yelin.
\newblock Cooperative resonances in light scattering from two-dimensional
  atomic arrays.
\newblock {\em Physical review letters}, 118(11):113601, 2017.

\bibitem{ref29}
MT~Manzoni, M~Moreno-Cardoner, A~Asenjo-Garcia, James~V Porto, Alexey~V
  Gorshkov, and DE~Chang.
\newblock Optimization of photon storage fidelity in ordered atomic arrays.
\newblock {\em New journal of physics}, 20(8):083048, 2018.

\bibitem{ref30}
David Plankensteiner, Christian Sommer, Helmut Ritsch, and Claudiu Genes.
\newblock Cavity antiresonance spectroscopy of dipole coupled subradiant
  arrays.
\newblock {\em Physical review letters}, 119(9):093601, 2017.

\bibitem{ref31}
Ana Asenjo-Garcia, M~Moreno-Cardoner, Andreas Albrecht, HJ~Kimble, and
  Darrick~E Chang.
\newblock Exponential improvement in photon storage fidelities using
  subradiance and “selective radiance” in atomic arrays.
\newblock {\em Physical Review X}, 7(3):031024, 2017.

\bibitem{ref32}
A~Grankin, PO~Guimond, DV~Vasilyev, B~Vermersch, and P~Zoller.
\newblock Free-space photonic quantum link and chiral quantum optics.
\newblock {\em Physical Review A}, 98(4):043825, 2018.

\bibitem{ref37}
Ephraim Shahmoon, Mikhail~D Lukin, and Susanne~F Yelin.
\newblock Quantum optomechanics of a two-dimensional atomic array.
\newblock {\em Physical Review A}, 101(6):063833, 2020.

\bibitem{ref38}
CD~Parmee and Janne Ruostekoski.
\newblock Bistable optical transmission through arrays of atoms in free space.
\newblock {\em Physical Review A}, 103(3):033706, 2021.

\bibitem{ref40}
Katharina Brechtelsbauer and Daniel Malz.
\newblock Quantum simulation with fully coherent dipole-dipole interactions
  mediated by three-dimensional subwavelength atomic arrays.
\newblock {\em Physical Review A}, 104(1):013701, 2021.

\bibitem{ref18}
Rivka Bekenstein, Igor Pikovski, Hannes Pichler, Ephraim Shahmoon, Susanne~F
  Yelin, and Mikhail~D Lukin.
\newblock Quantum metasurfaces with atom arrays.
\newblock {\em Nature Physics}, 16(6):676--681, 2020.

\bibitem{ref41}
Mariona Moreno-Cardoner, Daniel Goncalves, and Darrick~E Chang.
\newblock Quantum nonlinear optics based on two-dimensional rydberg atom
  arrays.
\newblock {\em Physical Review Letters}, 127(26):263602, 2021.

\bibitem{ref42}
Zhi-Yuan Wei, Daniel Malz, Alejandro Gonz{\'a}lez-Tudela, and J~Ignacio Cirac.
\newblock Generation of photonic matrix product states with rydberg atomic
  arrays.
\newblock {\em Physical Review Research}, 3(2):023021, 2021.

\bibitem{ref43}
David Fern{\'a}ndez-Fern{\'a}ndez and Alejandro Gonz{\'a}lez-Tudela.
\newblock Tunable directional emission and collective dissipation with quantum
  metasurfaces.
\newblock {\em Physical Review Letters}, 128(11):113601, 2022.

\bibitem{ref44}
Simon~Panyella Pedersen, Lida Zhang, Thomas Pohl, et~al.
\newblock Quantum nonlinear metasurfaces from dual arrays of ultracold atoms.
\newblock {\em Physical Review Research}, 5(1):L012047, 2023.

\bibitem{ref45}
Lida Zhang, Valentin Walther, Klaus M{\o}lmer, and Thomas Pohl.
\newblock Photon-photon interactions in rydberg-atom arrays.
\newblock {\em Quantum}, 6:674, 2022.

\bibitem{ref46}
Kritsana Srakaew, Pascal Weckesser, Simon Hollerith, David Wei, Daniel Adler,
  Immanuel Bloch, and Johannes Zeiher.
\newblock A subwavelength atomic array switched by a single rydberg atom.
\newblock {\em Nature Physics}, pages 1--6, 2023.

\bibitem{ref47}
Jun Rui, David Wei, Antonio Rubio-Abadal, Simon Hollerith, Johannes Zeiher,
  Dan~M Stamper-Kurn, Christian Gross, and Immanuel Bloch.
\newblock A subradiant optical mirror formed by a single structured atomic
  layer.
\newblock {\em Nature}, 583(7816):369--374, 2020.

\bibitem{ref26}
Roni Ben-Maimon, Yakov Solomons, and Ephraim Shahmoon.
\newblock Dissipative transfer of quantum correlations from light to atomic
  arrays.
\newblock {\em arXiv preprint arXiv:2311.03898}, 2023.

\bibitem{ref48}
Yakov Solomons and Ephraim Shahmoon.
\newblock Multichannel waveguide qed with atomic arrays in free space.
\newblock {\em Physical Review A}, 107(3):033709, 2023.

\bibitem{latt1}
Yeelai Chew, Takafumi Tomita, Tirumalasetty~Panduranga Mahesh, Seiji Sugawa,
  Sylvain de~L{\'e}s{\'e}leuc, and Kenji Ohmori.
\newblock Ultrafast energy exchange between two single rydberg atoms on a
  nanosecond timescale.
\newblock {\em Nature Photonics}, 16(10):724--729, 2022.

\bibitem{latt2}
Keisuke Nishimura, Hiroto Sakai, Takafumi Tomita, Sylvain de~L{\'e}s{\'e}leuc,
  and Taro Ando.
\newblock " super-resolution" holographic optical tweezers array.
\newblock {\em arXiv preprint arXiv:2411.03564}, 2024.

\bibitem{ref77}
Quantum error correction below the surface code threshold.
\newblock {\em Nature}, 638(8052):920--926, 2025.

\bibitem{ref78}
Daniel Hothem, Jordan Hines, Charles Baldwin, Dan Gresh, Robin Blume-Kohout,
  and Timothy Proctor.
\newblock Measuring error rates of mid-circuit measurements.
\newblock {\em Nature Communications}, 16(1):5761, 2025.

\bibitem{ref69}
Emma Deist, Yue-Hui Lu, Jacquelyn Ho, Mary~Kate Pasha, Johannes Zeiher, Zhenjie
  Yan, and Dan~M Stamper-Kurn.
\newblock Mid-circuit cavity measurement in a neutral atom array.
\newblock {\em Physical Review Letters}, 129(20):203602, 2022.

\bibitem{ref71}
Brandon Grinkemeyer, Elmer Guardado-Sanchez, Ivana Dimitrova, Danilo
  Shchepanovich, G~Eirini Mandopoulou, Johannes Borregaard, Vladan Vuleti{\'c},
  and Mikhail~D Lukin.
\newblock Error-detected quantum operations with neutral atoms mediated by an
  optical cavity.
\newblock {\em Science}, 387(6740):1301--1305, 2025.

\bibitem{ref66}
Trent~M Graham, Linipun Phuttitarn, Ravikumar Chinnarasu, Yunheung Song, Cody
  Poole, Kais Jooya, Jacob Scott, Abraham Scott, Patrick Eichler, and Mark
  Saffman.
\newblock Midcircuit measurements on a single-species neutral alkali atom
  quantum processor.
\newblock {\em Physical Review X}, 13(4):041051, 2023.

\bibitem{ref67}
Joanna~W Lis, Aruku Senoo, William~F McGrew, Felix R{\"o}nchen, Alec Jenkins,
  and Adam~M Kaufman.
\newblock Midcircuit operations using the omg architecture in neutral atom
  arrays.
\newblock {\em Physical Review X}, 13(4):041035, 2023.

\bibitem{ref68}
Kevin Singh, Conor~E Bradley, Shraddha Anand, Vikram Ramesh, Ryan White, and
  Hannes Bernien.
\newblock Mid-circuit correction of correlated phase errors using an array of
  spectator qubits.
\newblock {\em Science}, 380(6651):1265--1269, 2023.

\bibitem{ref74}
David Petrosyan and Klaus M{\o}lmer.
\newblock Deterministic free-space source of single photons using rydberg
  atoms.
\newblock {\em Physical Review Letters}, 121(12):123605, 2018.

\bibitem{ref72}
Hannah~J Manetsch, Gyohei Nomura, Elie Bataille, Kon~H Leung, Xudong Lv, and
  Manuel Endres.
\newblock A tweezer array with 6100 highly coherent atomic qubits.
\newblock {\em arXiv preprint arXiv:2403.12021}, 2024.

\bibitem{ref73}
Gr{\'e}goire Pichard, Desiree Lim, {\'E}tienne Bloch, Julien Vaneecloo, Lilian
  Bourachot, Gert-Jan Both, Guillaume M{\'e}riaux, Sylvain Dutartre, Richard
  Hostein, Julien Paris, et~al.
\newblock Rearrangement of individual atoms in a 2000-site optical-tweezer
  array at cryogenic temperatures.
\newblock {\em Physical Review Applied}, 22(2):024073, 2024.

\bibitem{ref75}
Inbal Friedler, David Petrosyan, Michael Fleischhauer, and Gershon Kurizki.
\newblock Long-range interactions and entanglement of slow single-photon
  pulses.
\newblock {\em Physical Review A—Atomic, Molecular, and Optical Physics},
  72(4):043803, 2005.

\bibitem{ref76}
Ofer Firstenberg, Charles~S Adams, and Sebastian Hofferberth.
\newblock Nonlinear quantum optics mediated by rydberg interactions.
\newblock {\em Journal of Physics B: Atomic, Molecular and Optical Physics},
  49(15):152003, 2016.

\end{thebibliography}
\end{document}